\documentclass[paper]{revtex4}
\usepackage{amsmath}

\usepackage{graphicx}
\usepackage{amssymb}
\usepackage{color}
\usepackage{enumerate}
\begin{document}

\title{On the occurrence of gauge-dependent secularities in nonlinear gravitational waves}
\author{Fabio Briscese}\email{fabio.briscese@northumbria.ac.uk, briscese.phys@gmail.com}
\affiliation{Department of Mathematics, Physics and Electrical
Engineering, Northumbria University, Newcastle City Campus, NE1
8ST, Newcastle upon Tyne, UK}

\affiliation{Istituto Nazionale di Alta Matematica Francesco
Severi, Gruppo Nazionale di Fisica Matematica, Citt\`a
Universitaria, P. le A. Moro 5, 00185 Rome, EU}

\author{Paolo Maria Santini}
\email{paolo.santini@roma1.infn.it} \affiliation{Dipartimento di
Fisica, Universit$\grave{a}$ di Roma "La Sapienza", 00185 Roma,
EU} \affiliation{Istituto Nazionale di Fisica Nucleare, Sezione di
Roma, EU.}

\begin{abstract}

We study the plane (not necessarily monochromatic) gravitational
waves at nonlinear quadratic order on a flat background in
vacuum. We show that, in the harmonic gauge, the nonlinear waves
are unstable. We argue that, at this order,  this instability can not be eliminated
by means of a multiscale approach, i.e. introducing suitable long
variables, as it is often the case when secularities appear in
a perturbative scheme. However, this is a non-physical and
gauge-dependent effect that disappears in a suitable system of
coordinates. In facts, we show that in a specific gauge such
instability does not occur, and that it is possible to solve
exactly the second order nonlinear equations of gravitational
waves. Incidentally, we note that this gauge coincides with the
one used by Belinski and Zakharov to find exact solitonic
solutions of Einstein's equations, that is to an exactly
integrable case, and this fact makes our second order nonlinear
solutions less interesting. However, the important warning is that
one must be aware of the existence of the instability reported in
this paper, when studying nonlinear gravitational waves in the
harmonic gauge.

\end{abstract}

\maketitle

\section{Introduction}

The LIGO-VIRGO collaboration  has recently detected two events,
GW150914 \cite{GW150914} and GW151226 \cite{GW151226}, relative to
the coalescence and merger of binary systems of black holes, and
the formation of a final black hole. This exceptional achievement
has finally directly proven the existence of gravitational waves
\footnote{It is worth to mention that the existence of
gravitational waves has been indirectly proved by the observation
of the inspiral motion of binary systems, as the famous
Hulse-Taylor Binary Pulsar PSR B1913+16 \cite{Hulse}, which are
characterized by a loss of kinetic energy  due to the emission of
gravitational waves (see \cite{GW review} for a review). We
mention that gravitational waves are also important in cosmology.
For instance, measurements of the cosmic microwave background
(CMB) anisotropy \cite{Planck} aim to reveal the existence of
primordial gravitational waves produced during the big bang.},
which is one of the main predictions of general relativity, as
pointed out by Einstein in 1916 \cite{Einstein}. The modelling of
the gravitational signal produced in the merge of binary systems
has played a crucial role in the direct detection of gravitational
waves, since the comparison between the data and the theoretically
predicted signal allows to discriminate between different types of
sources, and to estimate the physical parameters involved, e.g.
the black holes masses, spins and distances. This has been
achieved combining two (quite different) approaches, which are
used together to construct a model of the gravitational signal  at
different space scales (for instance in the "near-zone" and
"far-zone" with respect to the source) and for different parts of
the waveform (inspiral phase, merger and ringdown). The first
approach consists in the numerical solution of the Einstein's
equations, see \cite{pretorius} and references therein for a
review of numerical relativity results, which gives an useful
description of the merger of the black holes after the insipiral
phase. The second approach makes use of a perturbative expansion
of the Einstein's equations. The two most popular realizations of
this approach are the Post-Newtonian (PN) expansion
\cite{blanchet}, and the Effective One Body (EOB) formalism
\cite{damour}, which are successful to model the inspiral phase,
when the two black holes are well separated.

In this paper we are interested in the stability of the
perturbative approach. We consider the Einstein's equations in
vacuum for small perturbations $h$ of the Minkowski metric at
second order $h^2$ in the perturbative expansion in powers of $h$,
and we show that plane waves are unstable in the harmonic gauge,
which is commonly used to study gravitational waves in both the PN
and EOB formalisms. We  discuss this instability, and show that it
can not be eliminated by means of a multi scale approach
\footnote{The occurrence of secular instabilities in perturbation
theory is often due to the use of a bad perturbative scheme, in
problems in which the solutions depend simultaneously on widely
different scales, so that in such cases the secularities can be
avoided by means of a multi-scale perturbative expansion.}. In
spite of this fact, the nature of such instability is not
physical, instead it is a feature of the  harmonic gauge, so it is
simply a gauge effect. This conclusion is based on the observation
that, in an appropriate reference system explicitly defined, the
solutions of the nonlinear quadratic perturbative equations are
stable. In facts, in this gauge the second order nonlinear
equations are easily integrated, and their explicit solution shows
that the quadratic nonlinearities do not change in any significant
respect the picture of the gravitational waves obtained from the
linearized Einstein's equations. However, the relevance of these
perturbative solutions is diminished by the fact that the
reference system in which they have been found, is the same used
in the famous result of Zakharov and Belinski on gravitational
solitons \cite{zakharov belinski}, so that this gauge choice
corresponds to a situation in which the full Einstein's equations
in vacuum (and in electro-vacuum) are integrable.

The relevant point here is the existence of such gauge-induced
instability of plane waves in vacuum in the harmonic reference
system. This fact should be a warning for those who want to study
nonlinear gravitational waves  using the harmonic gauge in various
contexts. In facts,  in some cases the choice of the harmonic
gauge might not be convenient, due to the secular instability of
the perturbation $h$. For instance, in the PN and EOB formalisms
which make use of the harmonic gauge on a flat background, this
instability might appear if one studies the evolution of
gravitational waves far from the source using the plane wave
approximation. We mention that an analysis of this instability for
more general wavefronts and for different backgrounds is currently
under study, and will be reported elsewhere.

A further interesting question is whether this kind of instability
is still present in alternative gravitational models. For
instance,  in \cite{Cusin} it has been reported an instability of
gravitational waves in the bigravity model, which might be due
essentially to the same mechanism discussed in this paper.
Furthermore, a study of this instability in the case of modified
gravity \cite{capozziello} and of nonlocal gravitational models
\cite{nonlocal} would be also interesting, but it goes beyond the
purposes of this paper and will be discussed separately.

In what follows, unless explicitly stated, we will use the
following notations: Greeks indices run from zero to three, i.e.
$\alpha,\beta,\gamma\ldots =0,1,2,3$, Latin indices run from 2 to
3 $i,j,k \ldots =2,3$, capital Latin indices run from 1 to 3
$A,B,C \ldots =1,2,3$, and underlined indices run from $0$ to $1$
$\underline{a},\underline{b},\underline{c} \ldots = 0,1$. Moreover
we rise and lower indices with the Minkowski metric with signature
$1,-1,-1,-1$. The Ricci tensor is defined as in \cite{Landau} as
$R_{ij} = \partial_k \Gamma^k_{ij} + \ldots$.

\section{Einstein's equations}\label{einstein equations}

In this section we will write the Einstein's equations at second order of perturbations.
It is well known \cite{Weinberg,Landau}  that the components
$G^0_{\,\,\,\beta}$ of the Einstein's tensor do not contain second
time derivatives $\ddot{g}_{\alpha\beta}$ of the metric tensor,
but they contain only the first time derivative $\dot{g}_{AB}$ and
no derivatives $\dot{g}_{0 \alpha}$, so that

\begin{equation}
G^0_{\,\,\,\beta} = G^0_{\,\,\,\beta}(g_{\alpha\beta},
\dot{g}_{AB}, \partial_C g_{\alpha\beta},\partial_{C}\partial_D
g_{\alpha\beta} ) .
\end{equation}
Moreover the derivatives $\ddot{g}_{0\alpha}$ do not appear at all
and one has

\begin{equation}
G^A_{\,\,\, \gamma} = G^A_{\,\,\,
\gamma}(g_{\alpha\beta},\dot{g}_{0\alpha}, \dot{g}_{AB},
\ddot{g}_{AB},\partial_C g_{\alpha\beta},\partial_C
\partial_D g_{\alpha\beta} ) .
\end{equation}
Therefore the equations $G^0_{\,\,\,\beta}=0$   are involutive: if
the equations $G^A_{\,\,\,B} = 0$  are verified at any time and if
$G^0_{\,\,\, \alpha} = 0$  at the initial time $t_0$, then
$G^0_{\,\,\, \alpha} = 0$ at any time $t$. Thus, the equations
$G^0_{\,\,\, \alpha} = 0$ can be viewed as a constraint for
initial data. As a consequence, instead of 10 evolutionary
equations for the 10 independent components of the metric tensor
one has 6 evolutionary equations + 4 constraints on the initial
data, which leaves 4 of the 10 components of the metric arbitrary:
this arbitrariness corresponds to the arbitrariness in the choice
of the reference system \footnote{We refer to this arbitrariness
as gauge freedom, and to the choice of the reference frame as
gauge fixing.}. These considerations will be helpful in what
follows, when we will discuss the perturbative equations.

Since we want to study the propagation of nonlinear gravitational
waves in vacuum, we consider small perturbations  $h$ of the flat
spacetime metric, that is  a metric tensor $g_{\mu\nu} =
\eta_{\mu\nu} + h_{\mu\nu}$ with $\eta_{\mu\nu}$ the Minkowski
tensor and $|h_{\mu\nu}| = O(\epsilon)$, with $0<\epsilon \ll 1$.
From this definition it is possible to expand the Ricci tensor and
the Ricci scalar as well as the Einstein's tensor in powers of
$h$, see Appendix \ref{Appendix Einstein}.

For the Ricci tensor one has

\begin{equation}\label{definition ricci tot}
R_{\mu\nu} = R^{(1)}_{\mu\nu} + R^{(2)}_{\mu\nu} + O(h^3)
\end{equation}
where $R^{(1)}_{\mu\nu}$ and $ R^{(2)}_{\mu\nu}$ are respectively
the linear and quadratic part of the Ricci tensor, and are given
by

\begin{equation}
\begin{array}{ll}\label{Riccilinear}
R^{(1)}_{\mu\nu} = \frac{1}{2} \left[- \Box h_{\mu\nu} +
\partial_\mu \partial_\alpha \psi^\alpha_\nu +
\partial_\nu \partial_\alpha \psi^\alpha_\mu \right]= \frac{1}{2} \left[- h_{\mu\nu \quad ,\alpha}^{\quad,\alpha}
- h^{\alpha}_{\,\,\, \alpha ,\mu \nu} + h^{\alpha}_{\,\,\, \nu
,\alpha \mu } + h^{\alpha}_{\,\,\, \mu ,\alpha \nu } \right]
\end{array}
\end{equation}
and

\begin{equation}
\begin{array}{ll}\label{Ricciquadratic}
R^{(2)}_{\mu\nu} = \frac{1}{2} \{- \psi^{\alpha \rho}_{\quad ,
\alpha} \left( h_{\mu\rho,\nu} + h_{\nu\rho,\mu}-h_{\mu\nu,\rho}
\right)  +  h^{\alpha \rho} \left( h_{\mu\nu,\alpha\rho} +
h_{\alpha\rho,\mu\nu} - h_{\nu\rho,\alpha\mu}-
h_{\mu\rho,\alpha\nu}
\right) + \\
\\
+ \frac{1}{2} \left[h^{\alpha\rho}_{\,\,\,\, ,\mu}
h_{\alpha\rho,\nu}  + \left( h_{\mu\alpha,\rho} -
h_{\mu\rho,\alpha}\right) \left( h_{\nu}^{\,\,\,\alpha,\rho} -
h_{\nu}^{\,\,\,\rho,\alpha}\right) \right] \},
\end{array}
\end{equation}
where we have defined

\begin{equation}\label{definition psi}
\psi^\alpha_{\,\,\, \beta} \equiv h^\alpha_{\,\,\, \beta}-
\frac{1}{2} h^\rho_{\,\,\, \rho} \delta^\alpha_{\,\,\, \beta}
\end{equation}
see for instance \cite{Weinberg}.  The Ricci scalar is given by

\begin{equation}\label{Ricci scalar}
R = \eta^{\alpha\beta}R^{(1)}_{\alpha\beta} + \eta^{\alpha\beta}
R^{(2)}_{\alpha\beta} - h^{\alpha\beta} R^{(1)}_{\alpha\beta}+
O(h^3)
\end{equation}
and the Einstein's tensor up to second order in $h$ is given by

\begin{equation}\label{Einstein tensor 1}
G^\alpha_{\,\,\,\beta} \equiv R^\alpha_{\,\,\,\beta}-\frac{1}{2}
\delta^{\alpha}_{\beta} R = G^{(1)\alpha}_{\quad\,\,\,\beta} +
G^{(2)\alpha}_{\quad\,\,\,\beta} + O(h^3)
\end{equation}
where $G^{(1)\alpha}_{\quad\,\,\,\beta}$ and $
G^{(2)\alpha}_{\quad\,\,\,\beta}$ are respectively the linear and
quadratic parts of the Einstein's tensor, and are given by

\begin{equation}\label{Einstein tensor 2}
G^{(1)\alpha}_{\quad\,\,\,\beta} \equiv
R^{(1)\alpha}_{\quad\,\,\,\beta}- \frac{1}{2}
\delta^\alpha_{\beta} R^{(1)\gamma}_{\quad\,\,\,\,\gamma}
\end{equation}
and

\begin{equation}\label{Einstein tensor 3}
G^{(2)\alpha}_{\quad\,\,\,\beta} \equiv
R^{(2)\alpha}_{\quad\,\,\,\beta} -
h^{\alpha\sigma}R^{(1)}_{\quad\sigma\beta} - \frac{1}{2}
\delta^\alpha_{\beta} \left( R^{(2)\gamma}_{\quad\,\,\,\,\gamma} -
h^{\mu\nu}R^{(1)}_{\quad\mu\nu} \right)
\end{equation}

In this work we limit our analysis to plane waves and, without
loss of generality, we consider waves travelling along the $x^1$
axis, so that

\begin{equation}\label{planewaveassumption}
h_{\alpha \beta}=  h_{\alpha \beta}(x^0,x^1).
\end{equation}
Moreover, we choose $h$ as

\begin{equation}\label{polarizedNONlinearwave}
h(x^0,x^1) = \left(\begin{array}{cccc}
  h_{00} & h_{10} & 0 & 0 \\
  h_{10} & h_{11} & 0 & 0 \\
  0 & 0 & h_{22} & h_{23} \\
  0 & 0 & h_{23} & h_{33} \\
\end{array}\right)
\end{equation}
We stress that the choice $h_{0i}=0$ corresponds to a gauge choice, and therefore it does not affect the generality of the solution, while setting $h_{1i}=0$ is in facts a constraint on the form of the gravitational wave.

The waveform (\ref{polarizedNONlinearwave}) is compatible with the
Einstein's equations, up to the order $h^2$. In facts, the
equations $R_{ i\underline{a}}  = R^{(1)}_{\quad i\underline{a}} +
R^{(2)}_{\quad i\underline{a}} = 0$ are identically satisfied for
$h_{i\underline{a}} =0$, see (\ref{R1ai}) and (\ref{R2ia}).
Therefore one has  that $G^{\underline{a}}_{\,\,\,\, i}=
g^{\underline{a} \alpha} \, G_{\alpha i} = \left(
\eta^{\underline{a}
\,\underline{b}}-h^{\underline{a}\,\underline{b}}\right)
\left(R_{\,\underline{b} i} - g_{\,\underline{b} i} R/2\right) =
\delta^{\underline{a}}_{\,\,\, i} R/2 =0$, where we have used
$h^{\underline{a}\, i}=0$ and $\left( \eta^{\underline{a}
\,\underline{b}}-h^{\underline{a}\,\underline{b}}\right)
g_{\,\underline{b} i} = \delta^{\underline{a}}{\,\,\, i}$. Thus,
the equations $G^{\underline{a}}_{\,\,\,\, i}= 0$ are identically
satisfied for $h_{\underline{a}\, i}=0$ and one is left with the
four dynamical equations $G^{i}_{\,\,\, j}=0$ and $G^{1}_{\,\,\,
1}=0$ plus the two constrains $G^{0}_{\,\,\, 0}=0$ and
$G^{0}_{\,\,\, 1}=0$ for the six nonzero variables $h_{ij}$ and
$h_{\underline{a}\,\underline{b}}$. Thus, the solution
(\ref{polarizedNONlinearwave}) still contains two arbitrary
functions, corresponding to the residual gauge freedom, that can
be used  to fix the  gauge. Henceforth we will always set to zero
the components $h_{i\underline{a}}$ of the metric.

\subsection{Einstein's equations at first order}\label{Einstein's equations at first order}

We start with the Einstein's equations in vacuum, at first linear perturbative order, and we refer the reader to the  Appendix \ref{Appendix Einstein} for explicit calculations. Let us
define the following variables which will be useful in the
following discussions

\begin{equation}\label{definition A B PSI}
\begin{array}{ll}
B\equiv h_{22} + h_{33} \, ; \qquad \psi \equiv h_{22}-h_{23} \, ; \qquad
A \equiv h_{00,11} + h_{11,00} - 2 h_{01,01}
\end{array}
\end{equation}
Using these definitions, the linearized Einstein's equations for
the metric (\ref{polarizedNONlinearwave}) read

\begin{equation}\label{linear einstein eq new}
\begin{array}{ll}
G^{(1)0}_{\quad \,\, \, 0} = \frac{1}{2} B_{,11}=0 ; \qquad
G^{(1)0}_{\quad \,\, \, 1}  = \frac{1}{2} B_{,01}=0 ; \qquad
G^{(1)1}_{\quad \,\, \, 1}= - \frac{1}{2} B_{,00}=0  ; \qquad
G^{(1)2}_{\quad \,\, \, 3}=  \frac{1}{2}  \Box h_{23}=0\\
\\
G^{(1)2}_{\quad \,\, \, 2}=  \frac{1}{2} \left( \Box h_{22} - A
-\Box B \right)=0 ; \qquad
G^{(1)3}_{\quad \,\, \, 3}=  \frac{1}{2} \left( \Box h_{33} - A
-\Box B  \right)=0 .
\end{array}
\end{equation}
It is convenient to write the Einstein's equations by means of
linear combinations of (\ref{linear einstein eq new}) as follows

\begin{equation}
\begin{array}{ll} \label{linearequations}
G^{(1)0}_{\quad \,\,\,\, 0} + G^{(1)1}_{\,\,\,\, \quad 1} =   -
\frac{1}{2} \Box B = 0 , \qquad G^{(1)2}_{\,\,\,\, \quad 2} + G^{(1)3}_{\,\,\,\,
\quad3}-(G^{(1)0}_{\,\,\,\, \quad 0} + G^{(1)1}_{\,\,\,\, \quad1})
=  - A = 0 ,
\\
\\
G^{(1)2}_{\quad \,\,\,\, 3}  =  \frac{1}{2} \Box h_{23} =0 ,
\qquad G^{(1)2}_{\quad \,\,\,\, 2} - G^{(1)3}_{\quad \,\,\,\, 3}  =
\frac{1}{2} \Box \psi =0 ,
\end{array}
\end{equation}
together with the two constrains

\begin{equation}\begin{array}{ll}\label{linearconstrains new}
G^{(1)0}_{\quad \,\,\,\, 0}   =  \frac{1}{2} B_{,11} = 0 , \qquad G^{(1)0}_{\quad \,\,\,\, 1}   =  \frac{1}{2} B_{,01} = 0
\end{array}
\end{equation}
From equations (\ref{linearequations}-\ref{linearconstrains new})
one has  $B_{,00}= B_{,01}= B_{,11} = 0$, so that $B  = c_0 x^0 +
c_1 x^1 + c$, where $c_0,c_1, c$ are constants. Since we limit our
attention to bounded nonconstant solutions, we set $c_0=c_1=c=0$,
thus we have $B = h_{ 22} + h_{ 33} =0$. Moreover, since $h_{23}$
and $\psi$ satisfy  the wave equation, the components $h_{ij}$ are
a linear wave, that is $h_{ij}(x^0\pm x^1)$.   Furthermore, one
can always perform an infinitesimal gauge transformation
\begin{equation}\label{gauge transformation 1}
x'^0 = x^0 +   \xi^0(x^0\pm x^1), \qquad x'^1 = x^1 + \xi^1 (x^0
\pm x^1) ,
\end{equation}
under which the components $h_{\underline{a}\, \underline{b}}$
transform as

\begin{equation}\label{h transformation 1}
h'_{00} = h_{00} - 2\xi_{0,0} + O(\epsilon^2);\qquad h'_{11} =
h_{11} - 2\xi_{1,1} + O(\epsilon^2);\qquad h'_{01} = h_{01} -
\xi_{0,1} - \xi_{1,0} + O(\epsilon^2) ,
\end{equation}
and choose $\xi^0$ and $\xi^1$ in such a way that in the new coordinates
one has $h'_{\underline{a}\,\underline{b}} \sim \epsilon^2$, so that the gravitational wave reduces to the two well know polarizations

\begin{equation}\label{polarized+linearwave}
h'_+ = \left(\begin{array}{cccc}
  0 & 0 & 0 & 0 \\
  0 & 0 & 0 & 0 \\
  0 & 0 & h'_{+} & 0 \\
  0 & 0 & 0 & -h'_{+} \\
\end{array}\right), \qquad h'_\times = \left(\begin{array}{cccc}
  0 & 0 & 0 & 0 \\
  0 & 0 & 0 & 0 \\
  0 & 0 & 0 & h'_{\times}\\
  0 & 0 & h'_{\times} & 0 \\
\end{array}\right)
\end{equation}
up to quadratic  $\sim \epsilon^2$ corrections, where $h'_{+}$ and $h'_{\times}$ are solutions of the wave equation  (see \cite{Weinberg,Landau} for review).

\subsection{Einstein's equations at second order }

The Einstein's equations in vacuum at second perturbative order are

\begin{equation}
G^{\alpha}_{\,\,\, \beta} = G^{(1)\alpha}_{\quad\,\,\, \beta} + G^{(2)\alpha}_{\quad\,\,\,
\beta} + O(h^3)= 0 .
\end{equation}
Using equations (\ref{Ricci scalar}-\ref{Einstein tensor 3}) and
rearranging the Einstein's equations in the same form  as in
(\ref{linearequations}-\ref{linearconstrains new}), one obtains
(see Appendix \ref{Appendix Einstein})  the four dynamical
equations for the metric perturbation
(\ref{polarizedNONlinearwave})

\begin{subequations}\label{Einstein equations at second order}

\begin{equation}\label{e00+11-2}
\begin{array}{ll}
G^{0}_{\,\,\, \, 0}+ G^{1}_{\,\,\, \, 1} = \frac{1}{2} \{- \Box B+
B_{,00}  h_{00} +B_{,11}  h_{11}  - 2 B_{,01} h_{01}  +
\frac{1}{2} B_{,0} \left(h_{00,0}+ h_{11,0}  - 2 h_{01,1} \right) + \\
\\
+ \frac{1}{2} B_{,1} \left(h_{00,1}+ h_{11,1}  - 2 h_{01,0} \right) -h_{22} \Box h_{22}-h_{33} \Box h_{33}-2 h_{23} \Box h_{23}\\
\\
+\frac{1}{2} \left[\left( \psi_{,1} \right)^2 - \left( \psi_{,0}
\right)^2\right] + 2 \left[ \left( h_{23,1} \right)^2 - \left(
h_{23,0} \right)^2 \right] \} = 0 ,
\end{array}
\end{equation}

\begin{equation}\label{Emix}
\begin{array}{ll}
G^{2}_{\,\,\, \, 2}+ G^{3}_{\,\,\, \, 3}  - \left(G^{0}_{\,\,\, \,
0}+ G^{1}_{\,\,\, \, 1}\right) =\\
\\
=\frac{1}{2} \{-2 A + (h_{00,1})^2 -(h_{11,0})^2  +
 (h_{23,0})^2 -(h_{23,1})^2  + \frac{1}{4}
\left[(B_{,1})^2 -(B_{,0})^2 +(\psi_{,0})^2 -(\psi_{,1})^2
\right]+\\
\\
+ 2 \left(h_{00}-h_{11} \right) A  + h_{00,0} \left(h_{11,0}-2
h_{01,1} \right)- h_{11,1} \left(h_{00,1}-2 h_{01,0} \right)\} = 0 ,
\end{array}
\end{equation}

\begin{equation}\label{E23}
\begin{array}{ll}
G^{2}_{\,\,\, \, 3} = \frac{1}{2} \{\Box h_{23} -h_{00} h_{23,00}
- h_{11} h_{23,11} + 2 h_{01} h_{23,01} + h_{22}
(h_{23,00}-h_{23,11}) + h_{23} (h_{33,00}-h_{33,11}) +\\
\\
+\frac{1}{2} \left[ -h_{23,0} \left[\left(h_{00}+h_{11} -B
\right)_{,0} -2 h_{01,1} \right]  -h_{23,1}
\left[\left(h_{00}+h_{11} + B \right)_{,1} -2 h_{01,0}
\right]\right]\}=0 ,
\end{array}
\end{equation}

\begin{equation}\label{e22233}
\begin{array}{ll}
G^{2}_{ \,\,\, \, 2}-G^{3}_{\,\,\, \, 3} =\frac{1}{2} \{\Box
\left(h_{22} - h_{33} \right) + h_{00} \left(h_{33}- h_{22}
\right)_{,00} + 2 h_{01} \left(h_{22}-h_{33} \right)_{,01}
+ h_{11} \left(h_{33 } - h_{22}\right)_{,11} +\\
\\
-\frac{1}{2} \left( h_{22} - h_{33} \right)_{,0} \left(h_{00,0} +
h_{11,0} - B_{,0} -2 h_{01,1} \right) -\frac{1}{2} \left( h_{22} -
h_{33} \right)_{,1} \left(h_{00,1} + h_{11,1} + B_{,1} -2 h_{01,0}
\right)\\
\\
+ h_{22} \Box h_{22} -h_{33} \Box h_{33} \} = 0 ,
\end{array}
\end{equation}
and the two constrains on initial data

\begin{equation}\label{E00}
\begin{array}{ll}
G^{0}_{\,\,\, \, 0} = \frac{1}{2} \{ B_{,11}+ h_{11} B_{,11} -
h_{01} B_{,01}  + \frac{1}{2} B_{,0} \left(h_{11,0}  -2
h_{01,1}\right) +
\frac{1}{2} B_{,1} h_{11,1} + \\
\\
+\frac{1}{2} \left[(h_{22,1})^2 - (h_{22,0})^2 +3(h_{23,1})^2 -
(h_{23,0})^2 + h_{22,0} B_{,0}+ h_{33,1} (h_{33,1}-h_{22,1})
\right]+ \\
\\
+h_{22} h_{22,11}+h_{33} h_{33,11}+2 h_{23} h_{23,11} \} =0 ,
\end{array}
\end{equation}

\begin{equation}\label{E01}
\begin{array}{ll}
G^{0}_{\,\,\, \, 1} = \frac{1}{2} \{ B_{,01} +  h_{01} B_{,11} -
h_{00}
B_{,01} + \frac{1}{2} h_{11,0}  B_{,1}- \frac{1}{2} h_{00,1} B_{,0} +\\
\\
+\frac{1}{2} \left[h_{22,0}h_{22,1} +h_{33,0}h_{33,1} +
2h_{23,0}h_{23,1} \right]+ h_{22}h_{22,01} +h_{33}h_{33,01} +
2h_{23}h_{23,01} \}=0 .
\end{array}
\end{equation}

\end{subequations}

Let us discuss the properties of the solutions of the system
(\ref{Einstein equations at second order}). First of all note that
(\ref{e00+11-2}, \ref{E00},\ref{E01}) imply that $B_{,00}\sim
B_{,01}\sim B_{,11} \sim h^2$, so that $B = c + c_0 x^0  + c_1 x^1
+ O(h^2)$.  Again, since we are interested in bounded nonconstant
waves,  we choose $c=c_1 = c_2 = 0$, and therefore $B \sim h^2$.

Also note that $\Box h_{23}\sim \Box \psi \sim A \sim h^2$. That
implies that, since we are doing an analysis at second quadratic
order $h^2$, we must neglect all the terms $B h \sim h_{23} \,
\Box h \sim A h \ldots \sim h^3$ in the system  (\ref{Einstein
equations at second order}), so that the dynamical equations
reduce to the simplified form

\begin{subequations}\label{eeq}

\begin{equation}\label{eeq3}
\begin{array}{ll}
\Box h_{23} = h_{00} h_{23,00} + h_{11} h_{23,11} - 2 h_{01}
h_{23,01} +\frac{1}{2} h_{23,0} \left(h_{00,0}+h_{11,0}  -2
h_{01,1} \right) +\\
\\  +\frac{1}{2}  h_{23,1} \left(h_{00,1}+h_{11,1} -2 h_{01,0} \right) ,
\end{array}
\end{equation}

\begin{equation}\label{eeq4}
\begin{array}{ll}
\Box \psi = h_{00} \psi_{,00} + h_{11} \psi_{,11} - 2 h_{01}
\psi_{,01} +\frac{1}{2} \psi_{,0} \left(h_{00,0} + h_{11,0}  -2 h_{01,1} \right)+\\
\\ +\frac{1}{2} \psi_{,1} \left(h_{00,1} + h_{11,1}
 -2 h_{01,0} \right) ,
\end{array}
\end{equation}

\begin{equation}\label{eeq5}
\begin{array}{ll}
\Box B = \frac{1}{2} \left[\left( \psi_{,1} \right)^2 - \left(
\psi_{,0} \right)^2\right] + 2 \left[ \left( h_{23,1} \right)^2 -
\left( h_{23,0} \right)^2 \right] ,
\end{array}
\end{equation}

\begin{equation}\label{eeq6}
\begin{array}{ll}
 A = \frac{1}{2} \left[ (h_{00,1})^2 -(h_{11,0})^2  + (h_{23,0})^2 -(h_{23,1})^2 \right]  +
\frac{1}{8} \left[(\psi_{,0})^2 -(\psi_{,1})^2
\right]+\\
\\
+\frac{1}{2} h_{00,0} \left(h_{11,0}-2 h_{01,1} \right)-
\frac{1}{2} h_{11,1} \left(h_{00,1}-2 h_{01,0} \right) .
\end{array}
\end{equation}
In the same way,  the two constraints on the initial data reduce to

\begin{equation}\label{eeq1}
\begin{array}{ll}
B_{,11} +\frac{1}{2} \left[(h_{22,1})^2 - (h_{22,0})^2
+3(h_{23,1})^2 - (h_{23,0})^2 + h_{33,1} (h_{33,1}-h_{22,1})
\right] +\\
\\
+h_{22} h_{22,11}+h_{33} h_{33,11}+2 h_{23} h_{23,11} =0 ,
\end{array}
\end{equation}

\begin{equation}\label{eeq2}
\begin{array}{ll}
B_{,01} +\frac{1}{2} \left[h_{22,0}h_{22,1} +h_{33,0}h_{33,1} +
2h_{23,0}h_{23,1} \right]+ h_{22}h_{22,01} +h_{33}h_{33,01} +
2h_{23}h_{23,01} =0 .
\end{array}
\end{equation}
Note that the constraints (\ref{eeq1},\ref{eeq2}) depend only on
the components $h_{ij}$, which are the physical degrees of freedom
of the gravitational wave.

\end{subequations}

At this point we must fix the gauge, in order to discuss the
properties  of the solutions of the second order Einstein's
equations in the corresponding reference system. In facts, as
discussed in section \ref{einstein equations}, due to the residual
gauge invariance, the waveform (\ref{polarizedNONlinearwave})
still contains two arbitrary functions, that can be used to impose
two constraints on the components of the perturbations $h_{\alpha
\beta}$.

\section{Occurrence of the secularity in the harmonic
gauge}\label{section harmonic gauge}

In this section we show that, in the harmonic gauge, the plane waves of the form
(\ref{polarized+linearwave}) are unstable. The harmonic gauge is defined by the four conditions $\Gamma^\alpha_{\,\,\, \beta \gamma} g^{ \beta \gamma} \equiv \Gamma^\alpha = 0$, and it is
commonly used to study the generation and propagation of gravitational waves, e.g. in the PN and EOB formalisms \cite{blanchet,damour}.  Expanding  the equations $\Gamma^\lambda = 0$ at second order in $h$ one has

\begin{equation}\label{harmonicgaugecondition}
\Gamma^\lambda = \psi^{\lambda \alpha}_{\quad ,\alpha} -
h^{\lambda \sigma}\psi^{\alpha}_{\quad \sigma,\alpha} -
 h^{\alpha \beta} \left(h^{\lambda}_{\,\,\,\, \alpha,
\beta} - \frac{1}{2} h_{\alpha \beta}^{\quad ,\lambda} \right) +
O(h^3) = 0 ,
\end{equation}
which gives

\begin{equation}\label{harmonicgaugeconditionpsi}
\psi^{\lambda \alpha}_{\quad ,\alpha} = h^{\lambda
\sigma}\psi^{\alpha}_{\quad \sigma,\alpha} +
 h^{\alpha \beta} \left(h^{\lambda}_{\,\,\,\, \alpha,
\beta} - \frac{1}{2} h_{\alpha \beta}^{\quad ,\lambda} \right)
\sim h^2 .
\end{equation}
Since the perturbation $h_{\alpha \beta}$ is in the form
(\ref{polarizedNONlinearwave}), equation
(\ref{harmonicgaugeconditionpsi}) gives

\begin{subequations}
\begin{equation}\label{psidivergence1}
\partial_\alpha \psi^{\alpha 0} = \partial_\alpha \psi^{\alpha}_{\,\,\,\, 0} =
\left( \frac{h_{00} + h_{11} + h_{22} + h_{33}}{2}   \right)_{,0}
- h_{10,1} \qquad ,
\end{equation}

\begin{equation}\label{psidivergence2}
\partial_\alpha \psi^{\alpha 1} = - \partial_\alpha \psi^{\alpha}_{\,\,\,\, 1} =
\left( \frac{h_{00} + h_{11} - h_{22} - h_{33}}{2}   \right)_{,1}
- h_{10,0}  \qquad ,
\end{equation}

\begin{equation}\label{psidivergence3}
\partial_\alpha \psi^{\alpha}_{\,\,\,\, i} =  h^{\underline{a}}_{\,\,\,\, i,\underline{a}} = 0 ; \qquad i = 2,3 ; \qquad \underline{a} = 0,1 .
\end{equation}
\end{subequations}
Note that equations (\ref{psidivergence3}), corresponding to the
conditions  $\Gamma^i = 0$, are identically satisfied since
$h_{\underline{a} i} =0$ for   (\ref{polarizedNONlinearwave}).
Moreover, equations (\ref{psidivergence1}-\ref{psidivergence2}),
which correspond to the conditions $\Gamma^{\underline{a}} = 0$,
take the form

\begin{equation}\label{gauge2}
h_{01,1} - \left( \frac{h_{00} + h_{11} }{2} \right)_{,0} =
\frac{1}{2} B_{,0}
-h^{\underline{c}\,\underline{d}}h_{0\underline{c},\underline{d}}+\frac{1}{2}
h^{\underline{c}\,\underline{d}} h_{\underline{c}\,\underline{d},
0} +\frac{1}{2} h^{ij} h_{ij, 0} \, ,
\end{equation}
and

\begin{equation}\label{gauge3}
h_{01,0} - \left( \frac{h_{00} + h_{11} }{2} \right)_{,1} =  -
\frac{1}{2} B_{,1}
+h^{\underline{c}\,\underline{d}}h_{1\underline{c},\underline{d}}-\frac{1}{2}
h^{\underline{c}\,\underline{d}} h_{\underline{c}\,\underline{d}
,1} -\frac{1}{2} h^{ij} h_{ij, 1} \, ,
\end{equation}
Note that the right hand sides  of (\ref{gauge2}-\ref{gauge3})  are of
order $h^2$ since, as discussed in the previous section, we have
$B\sim h^2$.

For our purposes, it is more convenient to write the Einstein's equations in terms of the Ricci
tensor in the form $R_{\mu\nu} = 0$, that, through equations (\ref{definition ricci tot}-\ref{Ricciquadratic}), gives at the second perturbative order in $h$

\begin{equation}\label{einstein equations 2}
\Box h_{\mu\nu} = \partial_\mu \partial_\alpha
\psi^{\alpha}_{\,\,\, \nu} +\partial_\nu \partial_\alpha
\psi^{\alpha}_{\,\,\, \mu} + 2 R^{(2)}_{\mu \nu} .
\end{equation}
Assuming again that $h_{i \underline{a}} = 0$ and using equations
(\ref{R2ij}) and (\ref{R2ab}), equation (\ref{einstein equations
2}) gives up to quadratic terms

\begin{subequations}\label{Einstein equations at second order 2}

\begin{equation}\label{Einstein equations at second order 2 ij}
\Box h_{ij} = h^{\underline{a}\,\underline{b}}
h_{ij,\underline{a}\,\underline{b}}
\end{equation}

\begin{equation}\label{Einstein equations at second order 2 ab}
\Box h_{\underline{a}\,\underline{b}} = -\frac{1}{2}
h^{ij}_{\,\,\, ,\underline{a}} \,\, h_{ij, \underline{b}} +
M(h)_{\underline{a}\,\underline{b}}
\end{equation}
\end{subequations}
where

\begin{equation}
\begin{array}{ll}\label{m h}
M(h)_{\underline{a}\,\underline{b}} \equiv
\partial_{\underline{a}}\left[ h^{\underline{c}\, \underline{d}}
\left(h_{\underline{b}\,\underline{c}, \underline{d}}-\frac{1}{2}
h_{\underline{c}\, \underline{d}, \underline{b}} \right) \right]+
\partial_{\underline{b}}\left[ h^{\underline{c}\, \underline{d}}
\left(h_{\underline{a}\,\underline{c}, \underline{d}}-\frac{1}{2}
h_{\underline{c}\, \underline{d}, \underline{a}} \right)
\right]+\\
\\
+ h^{\underline{c}\,\underline{d}} \left(
h_{\underline{a}\,\underline{b},\underline{c}\,\underline{d}} +
h_{\underline{c}\,\underline{d},\underline{a}\,\underline{b}} -
h_{\underline{a}\,\underline{c},\underline{b}\,\underline{d}}-
h_{\underline{b}\,\underline{c},\underline{a}\,\underline{d}}
\right)+\\
\\
+\frac{1}{2} \left[h^{\underline{c}\,\underline{d}}_{\quad
,\underline{a}} h_{\underline{c}\,\underline{d},\underline{b}} +
\left( h_{\underline{a}\,\underline{c},\underline{d}} -
h_{\underline{a}\,\underline{d},\underline{c}}\right) \left(
h_{\underline{b}}^{\,\,\,\underline{c},\underline{d}} -
h_{\underline{b}}^{\,\,\,\underline{d},\underline{c}}\right)
\right]
\end{array}
\end{equation}

Equations (\ref{Einstein equations at second order 2}) imply that,
at the linear leading order $\sim \epsilon$, all the components
$h_{\alpha \beta}$ are solutions of the d'Alembert equation.
Therefore can write a plane nonlinear wave travelling along the
$x^1$ axis in the form

\begin{equation}\label{perturbative h}
h_{\alpha\beta} =  \epsilon \, h^{(1)}_{\alpha\beta}(x^0-x^1) +
\epsilon^2 \, h^{(2)}_{\alpha\beta}(x^0, x^1) + O(\epsilon^3) \, ,
\end{equation}
where for simplicity we choose waves travelling only in one
direction, and where $h^{(1)}_{\alpha\beta}, h^{(2)}_{\alpha\beta}
\sim 1 $. By means of (\ref{perturbative h}), the system
(\ref{Einstein equations at second order 2}) is identically
satisfied at linear $\epsilon$ order, while at quadratic order one
has

\begin{subequations}\label{Einstein equations at second order 3}

\begin{equation}\label{Einstein equations at second order 3 ij}
\Box h^{(2)}_{ij} = h^{(1)\underline{a}\,\underline{b}}
h^{(1)}_{ij,\underline{a}\,\underline{b}} =
\left(h^{(1)}_{00}+h^{(1)}_{11}+2 h^{(1)}_{01} \right)
 h^{(1)}_{ij,00} ,
\end{equation}

\begin{equation}\label{Einstein equations at second order 3 ab}
\Box h^{(2)}_{\underline{a}\,\underline{b}} =  -\frac{1}{2}
h^{(1)ij}_{\qquad \underline{a}} \,\, h^{(1)}_{ij, \underline{b}}
+ M(h^{(1)})_{\underline{a}\,\underline{b}}  =
(-1)^{\underline{a}+\underline{b}+1}   \, \kappa +
M(h^{(1)})_{\underline{a}\,\underline{b}} \, ,
\end{equation}
where we have defined

\end{subequations}

\begin{equation}\label{kappa}
\kappa \equiv \frac{1}{2} \left[ \left( h^{(1)}_{22,0} \right)^2
+\left( h^{(1)}_{33,0} \right)^2+ 2 \left( h^{(1)}_{23,0}
\right)^2\right]  .
\end{equation}

Note that, equations (\ref{gauge2}-\ref{gauge3},\ref{perturbative
h}) imply that $h^{(1)}_{00} + h^{(1)}_{11} + 2 h^{(1)}_{01} = c$
with $c$ constant, and since we are not interested in constant
solutions, we set $c=0$. Therefore, having $h^{(1)}_{00} +
h^{(1)}_{11} + 2 h^{(1)}_{01} = 0$,  equation (\ref{Einstein
equations at second order 3 ij}) reads $\Box h^{(2)}_{ij} = 0$, so
that also the second order components $h^{(2)}_{ij}(x^0-x^1)$ are
plane waves, and therefore they are stable.

In order to study the behavior or the second order components $
h^{(2)}_{\underline{a}\,\underline{b}}$, and without loss of
generality, in what follows we consider nonlinear gravitational
waves of the form

\begin{equation}\label{polarizedNONlinearwave 3}
h(x^0,x^1) = \epsilon \left(\begin{array}{cccc}
  0 & 0 & 0 & 0 \\
  0 & 0 & 0 & 0 \\
  0 & 0 & h^{(1)}_{22} & h^{(1)}_{23} \\
  0 & 0 & h^{(1)}_{23} & - h^{(1)}_{22} \\
\end{array}\right) + \epsilon^2 h^{(2)},
\end{equation}
In facts, it is always possible to use the coordinate transformation
(\ref{gauge transformation 1}) to cancel the contribution of the functions
$h^{(1)}_{\underline{a} \, \underline{b}}$  in
(\ref{perturbative h}), so that $h_{\underline{a} \,
\underline{b}} \sim \epsilon^2$ as long as $h^{(2)}_{\underline{a}
\, \underline{b}} \sim 1$
\footnote{This is exactly the same as in  the case of the linearized Einstein's
equations discussed in  section
\ref{Einstein's equations at first order}.}. Therefore, assuming that
the gravitational wave is in the form (\ref{polarizedNONlinearwave 3}),
which implies that $M(h^{(1)}) = 0$, the equation (\ref{Einstein equations at second order 3 ab})
for the components $h^{(2)}_{\underline{a}\,\underline{b}}$ becomes

\begin{equation}\label{Einstein equations at second order 4 ab}
\Box h^{(2)}_{\underline{a}\,\underline{b}}  =
(-1)^{\underline{a}+\underline{b}+1}   \, \kappa \, .
\end{equation}

Note that, equation(\ref{kappa})  implies that $\kappa >0 $ in
presence of the gravitational wave  (\ref{polarizedNONlinearwave
3}). Moreover, from (\ref{Einstein equations at second order 4
ab}) one immediately recognizes that $\kappa$ is a resonant
forcing for the components $h^{(2)}_{\underline{a} \,
\underline{b}}$, since form equation (\ref{kappa}) follows that
$\kappa$ is a function of $(x^0-x^1)$, and therefore a solution of
the equation $\Box \kappa = 0$. In conclusion, equation
(\ref{Einstein equations at second order 4 ab}) contains the
resonant forcing $\kappa$ and it is secular, so that the
components $h^{(2)}_{\underline{a} \, \underline{b}}$ grow with
time, thus the perturbative expansion in $h$ loses its asymptotic
character in the long time regime. That means that, in the
harmonic gauge, the linearized gravitational waves of the form
(\ref{polarized+linearwave}) are unstable, when second order
nonlinearities are considered.

Incidentally we also note that equation (\ref{Einstein equations
at second order 4 ab}) implies that

\begin{equation}\label{Einstein equations at second order h00h11h021}
\Box \left( h^{(2)}_{00} + h^{(2)}_{11} + 2 h^{(2)}_{01} \right) =
0 \, ,
\end{equation}
thus the quantity $h_{00} + h_{11} + 2 h_{01}$ is a solution of
the d'Alembert equation and it remains always finite.

Sometimes  in perturbation theory, the presence of secularities is
due to a wrong perturbative approach, in problems in which the
solutions depend simultaneously on widely different scales. In
such cases, the secularities can be eliminated introducing
suitable long variables; i.e., dealing with multiscale expansions.
However, our example is not curable in this way, and this seems to
be related to the fact that the instability found here is a
feature of the harmonic gauge, and the right way to treat it, is
to change the reference system.

Indeed, one can use a multiscale expansion
\begin{equation}
h_{ij}=\epsilon h^{(1)}_{ij}+\epsilon^2 h^{(2)}_{ij}+
O(\epsilon^{r}), \ \ \ h_{\underline{a}\,\underline{b}}=\epsilon
h^{(1)}_{\underline{a}\,\underline{b}}+ \epsilon^2
h^{(2)}_{\underline{a}\,\underline{b}}+ O(\epsilon^{r}),
\end{equation}
with $r  > 2$, introducing the slow variable $\tau$ as follows
\begin{equation}
\begin{array}{l}
h^{(1)}_{ij}=h^{(1)}_{ij}(\xi,\tau), \ \ \
h^{(1)}_{\underline{a}\,\underline{b}}=
h^{(1)}_{\underline{a}\,\underline{b}}(\xi,\tau), \\
\\
\xi=x^0-x^1, \ \ \ \tau=\epsilon^{N}x^0 .
\end{array}
\end{equation}
We do so, since we are interested in studying the solutions  in
the longtime regime $\xi=O(1)$ and $\tau = O(1)$. With this
assumptions, equation (\ref{Einstein equations at second order 2
ij}) reads

\begin{equation}\label{multiscale ab}
\epsilon \Box h^{(1)}_{ij} + \epsilon^2 \Box h^{(2)}_{ij} =
\epsilon^2 \left( h^{(1)}_{00} + h^{(1)}_{11} + 2
h^{(1)}_{01}\right)\partial_\xi^2 h^{(1)}_{ij} + O(\epsilon^3) + 2
\epsilon^{2+N} \left( h^{(1)}_{00} +
h^{(1)}_{01}\right)\partial_\xi
\partial_{\tau}h^{(1)}_{ij}.
\end{equation}
We have previously shown that the gauge conditions
(\ref{gauge2}-\ref{gauge3},\ref{perturbative h}) imply
$h^{(1)}_{00} + h^{(1)}_{11} + 2 h^{(1)}_{01} = O(\epsilon)$, thus
the first term in (\ref{multiscale ab}) is $O(\epsilon^3)$. The
second term in (\ref{multiscale ab}) is resonant, but we can avoid
it imposing $N>0$, so that (\ref{multiscale ab}) reduces to
(\ref{Einstein equations at second order 3 ij}). Let us stress
that, doing so, the $\tau$ dependence of the components
$h^{(1)}_{ij}(\xi,\tau)$, and consequently the $\tau$ dependence
of $\kappa(\xi,\tau)$, is not determined at order $\epsilon^2$. We also emphasize that
(\ref{Einstein equations at second order 3 ij}) does not need a
multiscale expansion, indeed the components $h_{ij}$ are not
secular, even in the standard perturbative expansion. Instead, the
appearance of the second resonant term in (\ref{multiscale ab}) is
related to the introduction of the slow variable $\tau$, so that
it can be avoided if $\tau$ is not introduced at all. However, let
us continue to use our multiscale approach, to show that it is not
useful to cure the secularities in the components
$h_{\underline{a}\underline{b}}$.

Let us consider the equation (\ref{Einstein equations at second
order 2 ab}), which at $O(\epsilon^2)$ reads

\begin{equation}
2\epsilon^{N+1}\partial_{\xi}\partial_{\tau}h^{(1)}_{\underline{a}\underline{b}}(\xi,\tau)+
\epsilon^2 \Box h^{(2)}_{\underline{a}\underline{b}}= \epsilon^2
\left[(-1)^{\underline{a}+\underline{b}+1}\kappa(\xi,\tau)+M_{\underline{a}\underline{b}}\left(h^{(1)}(\xi,\tau)\right)\right]
\,  .
\end{equation}
For the principle of maximum balance one has $N = 1$, and using
the fact that
$M_{\underline{a}\underline{b}}\left(h^{(1)}(\xi,\tau)\right)=0$
for the waveform (\ref{polarizedNONlinearwave 3}), one has

\begin{equation}
\Box
h^{(2)}_{\underline{a}\underline{b}}=-2\partial_{\xi}\partial_{\tau}h^{(1)}_{\underline{a}\underline{b}}(\xi,\tau)+
(-1)^{\underline{a}+\underline{b}+1}\kappa(\xi,\tau).
\end{equation}
Since the r.h.s. of this equation is a secular forcing for
$h^{(2)}_{\underline{a}\underline{b}}$, we should set it to $0$ in
order to avoid the secularity, also defining in that way the
$\tau$ dependence of $h^{(1)}_{\underline{a}\underline{b}}$
through the equation
\begin{equation}\label{multiscale ij}
2\partial_{\xi}\partial_{\tau}h^{(1)}_{\underline{a}\underline{b}}(\xi,\tau)=(-1)^{\underline{a}+\underline{b}+1}\kappa(\xi,\tau)
\, ,
\end{equation}
valid in the longtime regime $\xi = O(1)$ and $\tau= \epsilon x^0
= O(1)$. However, since the $\tau$ dependence of
$\kappa(\xi,\tau)$ is not defined by (\ref{multiscale ab}), it is
not possible to obtain the $\tau$ dependence of
$h^{(1)}_{\underline{a}\underline{b}}(\xi,\tau)$ through the
equation (\ref{multiscale ij}). Therefore, the multiscale method
does not fix the dependence of $h_{\mu\nu}$ on the slow  variable
$\tau$ at order $\epsilon^2$, and we conclude that, at this order,
the  secularity in the components $h_{\underline{a}\underline{b}}$
can not be eliminated by means of  a multiscale perturbative
expansion.

\section{Stability of the perturbative scheme in a different
gauge}

In order to prove that the instability discussed in the previous
section is gauge dependent, in what follows we show that, in an
opportune gauge, the evolution of gravitational waves at second
perturbative order is stable. Therefore, we come back to the
waveform (\ref{polarizedNONlinearwave}), which contains two
arbitrary functions, and we fix such functions imposing different
gauge conditions from (\ref{harmonicgaugecondition}), which
characterize the harmonic gauge. Since we want to exploit the
properties of the d'Alembert equation, we impose the following
gauge conditions

\begin{equation}\label{gauge1}
h_{00} = - h_{11}; \qquad h_{01} = 0
\end{equation}
which imply that

\begin{equation}
A = h_{00,11} + h_{11,00} - 2 h_{01,01}=  \Box h_{11}
\end{equation}
Using (\ref{gauge1}) and neglecting the terms $O(h^3)$, the
equations (\ref{eeq3}) and (\ref{eeq4}) reduce to the homogeneous
wave equations

\begin{equation}\label{eeeq1}
\begin{array}{ll}
\Box h_{23}   = 0, \qquad \Box \psi =  0 .
\end{array}
\end{equation}
Since $h_{23}$ and $\psi$ are plane waves solutions of the
d'Alembert equation, it is immediate to recognize that the right
hand side of equation (\ref{eeq5}) is null, so that also $B$ is
solution of the homogeneous wave equation

\begin{equation}\label{eeeq2}
\begin{array}{ll}
\Box B = 0 .
\end{array}
\end{equation}
Therefore the functions $h_{23}$, $B$ and $ \psi$ are plane waves
travelling along the $x^1$ axis

\begin{equation}\label{sol1}
\begin{array}{ll}
h_{23} = h^+_{23}(x^0-x^1) + h^-_{23}(x^0+x^1) \sim h \\
\\
\psi = \psi^+(x^0-x^1) + \psi^-(x^0+x^1) \sim h\\
\\
B = B^+(x^0-x^1) + B^-(x^0+x^1) \sim h^2 \, ,
\end{array}
\end{equation}
and are such that they satisfy the constraints
(\ref{eeq1},\ref{eeq2}) on their initial values. Moreover,
evaluating equation(\ref{eeq6}) over the solutions (\ref{sol1}) in
the gauge (\ref{gauge1}) one has

\begin{equation}\label{eeeqh00}
\begin{array}{ll}
\Box h_{11} =    \left(h_{11,1} \right)^2 - \left(h_{11,0}
\right)^2
\end{array}
\end{equation}
whose general solution  is given by

\begin{equation}\label{solh00}
\begin{array}{ll}
h_{11} =  \ln \left( 1 + \alpha(x^0-x^1) + \beta(x^0+x^1) \right)
\end{array}
\end{equation}
where $\alpha(x^0-x^1) \sim \epsilon$ and $ \beta(x^0+x^1)\sim
\epsilon$ so that $h_{11} \sim \epsilon$ (see Appendix
\ref{Appendix prototipe equation} for the solution of the equation
(\ref{eeeqh00}) ). We conclude that equations
(\ref{sol1},\ref{solh00}) gives the general solution of the
Einstein's equations in vacuum, at second order of perturbations
$h^2$, in the gauge (\ref{gauge1}).

At this point it is worth to notice that our gauge choice
(\ref{gauge1}) coincides with the one introduced by Zakharov and
Belinski  in their studies on gravitational solitons
\cite{zakharov belinski}, in which the metric tensor is in the
form

\begin{equation}
ds^2 = f(x^0,x^1) \left((dx^0)^2 - (dx^1)^2 \right) + g_{ij} dx^i
dx^j \, .
\end{equation}
Therefore, in the gauge (\ref{gauge1}), the Einstein's equations
can be resolved exactly with the inverse scattering technique, so
that our solution is just a perturbative result in an exactly
integrable context. However, here we were interested in showing
that the secularity discussed in section \ref{section harmonic
gauge}  is  gauge-dependent, and it is a feature of the harmonic
reference system.

To conclude this section, let us show that the solution
(\ref{sol1},\ref{solh00}) can be recast to coincide with
(\ref{polarized+linearwave}) at linear order $\sim \epsilon$.
Since $\alpha\sim \beta\sim \epsilon$, from  (\ref{solh00}) one
has

\begin{equation}\label{solh00-1}
\begin{array}{ll}
h_{00} = \alpha(x^0-x^1) + \beta(x^0+x^1) + O(\epsilon^2)
\end{array}
\end{equation}
Under an infinitesimal change of coordinates of the type $x'^0 =
x^0 +   \xi^0(x^0,x^1)$, $x'^1 = x^1 + \xi^1 (x^0,x^1)$, the
components $h_{\underline{a}\, \underline{b}}$   transform as

\begin{equation}\label{diffeo2}
h'_{00} = h_{00} - 2\xi_{0,0} + O(\epsilon^2);\qquad h'_{00} =
h_{11} - 2\xi_{1,1} + O(\epsilon^2);\qquad h'_{01} = h_{01} -
\xi_{0,1} - \xi_{1,0} + O(\epsilon^2) .
\end{equation}
Choosing $\xi_0$ and $\xi_1$ as

\begin{equation}\label{diffeo4}
\begin{array}{ll}
\xi_{0}= \xi^+(x^0-x^1) + \xi^-(x^0+x^1) , \qquad
\xi_{1} = \xi^+(x^0-x^1) - \xi^-(x^0+x^1)\\
\\
\xi^+(r) \equiv \frac{1}{2} \int^r  \alpha(r') \, dr' , \qquad
\xi^-(s) \equiv \frac{1}{2} \int^s  \beta(s') \, ds',
\end{array}
\end{equation}
from (\ref{diffeo2}) one has $ h'_{00} = O(\epsilon^2)$, and $
h'_{11} = O(\epsilon^2) $. Moreover, using (\ref{diffeo4}) and the
fact that $h_{01}= 0$, one has $h'_{01} = - \xi_{0,1}- \xi_{1,0} +
O(\epsilon^2) = O(\epsilon^2)$, therefore all the components
$h'_{\underline{a}\,\underline{b}}$ are of order $\sim
\epsilon^2$, and they can be always neglected at the dominant
order $\epsilon$. For the components $h_{ij}$, using (\ref{sol1})
one has

\begin{equation}\label{diffeo7}
\begin{array}{ll}
h'_{ij} = h^+_{ij}\left(x^0-x^1+ \xi^0-\xi^1 \right) +
h^-_{ij}\left(x^0+x^1+ \xi^0+\xi^1 \right)=\\
\\
= h^+_{ij}\left(x^0-x^1+  2 \,  \xi^+(x^0-x^1) \right)+
h^-_{ij}\left(x^0+x^1+  2 \,  \xi^-(x^0+x^1)\right) \sim \epsilon
\, ,
\end{array}
\end{equation}
which shows how the nonlinearity mildly changes the waveform of
the plane wave. In conclusion, in the appropriate gauge
(\ref{gauge1}), the nonlinear gravitational waves will be of the
usual form

\begin{equation}\label{polarizedNONlinearwave 2}
h(x^0,x^1) = \left(\begin{array}{cccc}
  0 & 0 & 0 & 0 \\
  0 & 0 & 0 & 0 \\
  0 & 0 & h_{22} & h_{23} \\
  0 & 0 & h_{23} & - h_{22} \\
\end{array}\right) + O(\epsilon^2)
\end{equation}
with $h_{22}$ and $h_{23}$ given by (\ref{diffeo7}).

\section{Conclusions}

In this paper we have studied the properties of nonlinear plane
(but non necessarily  monochromatic) gravitational waves. More
specifically, we have analyzed the evolution of small
perturbations $h$ of the Minkowski metric of the form
(\ref{planewaveassumption}-\ref{polarizedNONlinearwave}), at
quadratic $h^2$ perturbative level, in vacuum. We have shown that,
in the harmonic gauge, which is usually used to study
gravitational waves, the components (for a plane wave moving in
the direction $x^1$) $h_{00}$, $h_{11}$ and $h_{11}$ grow with
time, thus gravitational waves of the form
(\ref{polarized+linearwave}) are unstable. Therefore, in some
cases, the harmonic gauge might not be the best choice to study
the evolution of gravitational waves perturbatively. We mention
that the instability reported in this paper resembles that
appearing in \cite{aldrovndi}, where the authors have studied the
effect of quadratic nonlinearities on monochromatic gravitational
waves. However, in \cite{aldrovndi} the authors do not infer that
the growth of the components $h_{\underline{a}\underline{b}}$
represents a failure of the perturbative analysis, and they do not
provide a solution or interpretation of this fact.

Finally, we have argued that the instability found here, is
characteristic of the harmonic gauge, thus it is a gauge-dependent
feature. In facts, we have shown that, in the gauge (\ref{gauge1}),
the waveform
(\ref{planewaveassumption}-\ref{polarizedNONlinearwave}) is
stable, and the dynamic and the properties of nonlinear
gravitational waves are essentially the same of linear
gravitational waves, given by the two polarizations in
(\ref{polarized+linearwave}).

\appendix

\section{Ricci tensor} \label{Appendix Ricci}

In this appendix we calculate the Ricci tensor and the Ricci
scalar at linear and quadratic orders as defined in equations
(\ref{definition ricci tot}-\ref{Ricci scalar}), assuming that
$h_{\mu \nu}$ is a plane  wave travelling along the $x^1$ axis, of
the form (\ref{planewaveassumption}).

\subsection{First order contribution to the Ricci tensor}

For the linear contribution to the Ricci tensor
(\ref{Riccilinear}) one has

\begin{equation}\label{R100}
R^{(1)}_{00} = \frac{1}{2} \{h_{00,11} +h_{11,00} +
\left(h_{22}+h_{33} \right)_{,00}- 2 \, h_{10,10} \}
\end{equation}

\begin{equation}\label{R101}
R^{(1)}_{01} = \frac{1}{2} \left(h_{22}+h_{33} \right)_{,01}
\end{equation}

\begin{equation}\label{R111}
R^{(1)}_{11} = \frac{1}{2} \{- \left(h_{00,11} +h_{11,00}\right) +
\left(h_{22}+h_{33} \right)_{,11}+ 2 \, h_{10,10} \}
\end{equation}

Together with

\begin{equation}\label{R1ij}
R^{(1)}_{ij} = - \frac{1}{2}  \Box h_{ij}
\end{equation}

and

\begin{equation}\label{R1ai}
R^{(1)}_{\underline{a}\, i} = \frac{1}{2} \{ -\Box
h_{\underline{a}\, i} - h^{\underline{b}}_{\,\,\,
i,\underline{b}\,\underline{a}} \}
\end{equation}
Note that, if $h_{\underline{a}\, i}$ vanishes, the components
$R^{(1)}_{\underline{a}\, i}$ are identically zero. Therefore,
setting $h_{\underline{a}\, i} =0$ is allowed by the linearized
Einstein's equations.

The Ricci scalar at first order is

\begin{equation}\label{Ricci1}
R^{(1)\alpha}_{\quad\,\,\,\,\alpha}  = h_{11,00} + h_{00,11} - 2\,
h_{01,01 } + \left(  h_{ 22} + h_{ 33}\right)_{,00} - \left( h_{
22} + h_{ 33}\right)_{,11}
\end{equation}

\subsection{Second order contribution to the Ricci tensor}

Let us calculate the second order contribution to the Ricci tensor
as given in (\ref{Ricciquadratic})

\subsubsection{Components ${i\underline{a}}$} \label{Appendix Ricci ia}

Let us calculate the components $R^{(2)}_{\quad i\,\underline{a}}$
with $i=2,3$ and $\underline{a} = 0,1$. One has

\begin{equation}
\begin{array}{ll}
- \psi^{\alpha \rho}_{\quad , \alpha} \left(
h_{i\rho,\underline{a}} +
h_{\underline{a}\rho,i}-h_{i\underline{a},\rho} \right)   = -
\psi^{\underline{c}\, \underline{b}}_{\quad , \underline{c}}
\left( h_{i\underline{b},\underline{a}}
-h_{i\underline{a},\underline{b}} \right) -
 h^{\underline{b}\, j}_{\,\,\,\, ,\underline{b}} \, h_{ij,\underline{a}}\\
\\
h^{\alpha \rho} \left( h_{i\, \underline{a},\alpha\rho} +
h_{\alpha\rho,i\, \underline{a}} - h_{ \underline{a}\,\rho,\alpha
i}- h_{i\rho,\alpha \, \underline{a}} \right) =
h^{\underline{b}\,\underline{c}} \left( h_{i\, \underline{a},
\underline{b}\,\underline{c}}  - h_{i\, \underline{c},
\underline{b}\,\underline{a}} \right) -  h^{\underline{b} \, h}
h_{ih, \underline{b}\, \underline{a}} \\
\\
h^{\alpha\rho}_{\quad ,i} h_{\alpha\rho,\underline{a}}  =0 \\
\\
\frac{1}{2}\left( h_{i\alpha,\rho} - h_{i\rho,\alpha}\right)
\left( h_{\underline{a}}^{\,\,\,\alpha,\rho} -
h_{\underline{a}}^{\,\,\,\rho,\alpha}\right) = \frac{1}{2}\left(
h_{i \underline{c}, \underline{b}} - h_{i \underline{b},
\underline{c}}\right) \left( h_{\underline{a}}^{\,\,\,
\underline{c}, \underline{b}} -
h_{\underline{a}}^{\,\,\,\underline{b}, \underline{c}}\right) + 2
\, h_{i h, \underline{b}} \,  h_{\underline{a}}^{\,\,\, h,
\underline{b}}
\end{array}
\end{equation}

which gives

\begin{equation}\label{R2ia}
\begin{array}{ll}
R^{(2)}_{i\underline{a}} = \frac{1}{2} \{ - \psi^{ \underline{c}\,
\underline{b}}_{\quad , \underline{c}} \left(
h_{i\underline{b},\underline{a}} -h_{i\underline{a},\underline{b}}
\right) + h^{\underline{b}\,\underline{c}} \left( h_{i\,
\underline{a}, \underline{b}\,\underline{c}}  - h_{i\,
\underline{c},
\underline{b}\,\underline{a}} \right)  + \\
\\
 +\frac{1}{2}\left( h_{i \underline{c}, \underline{b}} -
h_{i \underline{b}, \underline{c}}\right) \left(
h_{\underline{a}}^{\,\,\, \underline{c}, \underline{b}} -
h_{\underline{a}}^{\,\,\,\underline{b}, \underline{c}}\right) + 2
\, h_{i h, \underline{b}} \,  h_{\underline{a}}^{\,\,\, h,
\underline{b}} - h^{\underline{b}\, j}_{\,\,\,\, ,\underline{b}}
\, h_{ij,\underline{a}} -  h^{\underline{b} \, h} h_{ih,
\underline{b}\, \underline{a}} \}
\end{array}
\end{equation}

Note that if the components $h_{i\underline{a}}$ are null, the
Einstein's equations for the components ${i\underline{a}}$, namely
$R^{(1)}_{\quad i\underline{a}} + R^{(2)}_{\quad i\underline{a}} =
0$ are identically satisfied (in facts $R^{(1)}_{\quad
i\underline{a}} = 0$ and $R^{(2)}_{\quad i\underline{a}} = 0$
separately), so that the choice $h_{i\underline{a}} =0$ is allowed
by the Einstein's equations. Therefore, henceforth we will keep
this assumption in our equations and we will always neglect the
components $h_{i\underline{a}}$ of the metric.

\subsubsection{Components ${i,j}$}

Let us calculate the components $R^{(2)}_{i,j}$ with $i,j=2,3$.
All the derivatives with respect to $x^i$ and $x^j$ will be zero,
so

\begin{equation}
\begin{array}{ll}
- \psi^{\alpha \rho}_{\quad , \alpha} \left( h_{i\rho,j} +
h_{j\rho,i}-h_{ij,\rho} \right)   =  \psi^{\underline{a}\,
\underline{b}}_{\quad , \underline{a}} \, h_{ij,\underline{b}} \\
\\
h^{\alpha \rho} \left( h_{ij,\alpha\rho} + h_{\alpha\rho,ij} -
h_{j\rho,\alpha i}- h_{i\rho,\alpha j} \right) =
h^{\underline{a}\,\underline{b}}
h_{ij,\underline{a}\,\underline{b}} \\
\\
h^{\alpha\rho}_{\quad ,i} h_{\alpha\rho,j}  =0 \\
\\
\frac{1}{2}\left( h_{i\alpha,\rho} - h_{i\rho,\alpha}\right)
\left( h_{j}^{\,\,\,\alpha,\rho} -
h_{j}^{\,\,\,\rho,\alpha}\right) = \frac{1}{2}\left( h_{i
\underline{a}, \underline{b}} - h_{i \underline{b},
\underline{a}}\right) \left( h_{j}^{\,\,\, \underline{a},
\underline{b}} - h_{j}^{\,\,\,\underline{b}, \underline{a}}\right)
+ h_{i h, \underline{a}} h_{j}^{\,\,\, h, \underline{a}}
\end{array}
\end{equation}

Therefore

\begin{equation}
R^{(2)}_{ij} = \frac{1}{2} \{  \psi^{\underline{a}\,
\underline{b}}_{\quad , \underline{a}} \, h_{ij,\underline{b}} +
h^{\underline{a}\,\underline{b}}
h_{ij,\underline{a}\,\underline{b}} + \frac{1}{2}\left( h_{i
\underline{a}, \underline{b}} - h_{i \underline{b},
\underline{a}}\right) \left( h_{j}^{\,\,\, \underline{a},
\underline{b}} - h_{j}^{\,\,\,\underline{b}, \underline{a}}\right)
+ h_{i h, \underline{a}} h_{j}^{\,\,\, h, \underline{a}}\}
\end{equation}

Under the condition $h_{i\underline{a}}=0$ one has

\begin{equation}\label{R2ij}
R^{(2)}_{ij} = \frac{1}{2} \{  \psi^{\underline{a}\,
\underline{b}}_{\quad , \underline{a}} \, h_{ij,\underline{b}} +
h^{\underline{a}\,\underline{b}}
h_{ij,\underline{a}\,\underline{b}} + h_{i h, \underline{a}}
h_{j}^{\,\,\, h, \underline{a}}\}
\end{equation}

\subsubsection{Components $\underline{a}\underline{b}$}

Let us calculate the components $R^{(2)}_{\quad \underline{a}\,
\underline{b}}$. One has

\begin{equation}
\begin{array}{ll}\label{Ricciquadraticab}
- \psi^{\alpha \rho}_{\quad , \alpha} \left(
h_{\underline{b}\,\rho,\underline{a}} +
h_{\underline{a}\,\rho,\underline{b}}-h_{\underline{b}\,\underline{a},\rho}
\right) = - \psi^{\underline{d}\, \underline{c}}_{\quad ,
\underline{d}} \left(
h_{\underline{b}\,\underline{c},\underline{a}}
+h_{\underline{a}\,\underline{c},\underline{b}}
-h_{\underline{b}\,\underline{a},\underline{c}} \right) \\
\\
h^{\alpha \rho} \left( h_{\underline{b}\,
\underline{a},\alpha\rho} + h_{\alpha\rho,\underline{b}\,
\underline{a}} - h_{ \underline{a}\,\rho,\alpha \, \underline{b}}-
h_{\underline{b}\,\rho,\alpha \, \underline{a}} \right) = h^{ij}
h_{ij,\underline{a}\,\underline{b}} +  h^{ \underline{c}\,
\underline{d}} \left( h_{\underline{b}\, \underline{a},
\underline{c}\,\underline{d}} +
h_{\underline{c}\,\underline{d},\underline{b}\, \underline{a}} -
h_{ \underline{a}\, \underline{c}, \underline{d} \,
\underline{b}}-
h_{\underline{b}\, \underline{c}, \underline{d} \, \underline{a}} \right) \\
\\
\frac{1}{2}h^{\alpha\rho}_{\quad ,\underline{b}}
h_{\alpha\rho,\underline{a}} = \frac{1}{2}h^{\underline{c}\,
\underline{d}}_{\quad ,\underline{b}} h_{\underline{c}\,
\underline{d},\underline{a}} +
\frac{1}{2}h^{ij}_{\quad ,\underline{b}} h_{ij,\underline{a}}\\
\\
\frac{1}{2}\left( h_{\underline{b}\,\alpha,\rho} -
h_{\underline{b}\,\rho,\alpha}\right) \left(
h_{\underline{a}}^{\,\,\,\alpha,\rho} -
h_{\underline{a}}^{\,\,\,\rho,\alpha}\right) = \frac{1}{2}\left(
h_{\underline{b}\, \underline{c}, \underline{d}} -
h_{\underline{b}\, \underline{d}, \underline{c}}\right) \left(
h_{\underline{a}}^{\,\,\, \underline{c}, \underline{d}} -
h_{\underline{a}}^{\,\,\,\underline{d}, \underline{c}}\right)
\end{array}
\end{equation}

so that

\begin{equation}\label{R2ab}
\begin{array}{ll}
R^{(2)}_{\underline{b}\, \underline{a}} = \frac{1}{2} \{ -
\psi^{\underline{d} \,\underline{c}}_{\quad , \underline{d}}
\left( h_{\underline{b}\,\underline{c},\underline{a}}
+h_{\underline{a}\,\underline{c},\underline{b}}
-h_{\underline{b}\,\underline{a},\underline{c}} \right) + h^{
\underline{c}\, \underline{d}} \left( h_{\underline{b}\,
\underline{a}, \underline{c}\,\underline{d}} +
h_{\underline{c}\,\underline{d},\underline{b}\, \underline{a}} -
h_{ \underline{a}\, \underline{c}, \underline{d} \,
\underline{b}}- h_{\underline{b}\, \underline{c}, \underline{d} \,
\underline{a}} \right) + \\
\\
\frac{1}{2}\left( h_{\underline{b}\, \underline{c}, \underline{d}}
- h_{\underline{b}\, \underline{d}, \underline{c}}\right) \left(
h_{\underline{a}}^{\,\,\, \underline{c}, \underline{d}} -
h_{\underline{a}}^{\,\,\,\underline{d}, \underline{c}}\right) +
\frac{1}{2}h^{\underline{c}\, \underline{d}}_{\quad
,\underline{b}} h_{\underline{c}\, \underline{d},\underline{a}} +
\frac{1}{2}h^{ij}_{\quad ,\underline{b}} h_{ij,\underline{a}} +
h^{ij} h_{ij,\underline{a}\,\underline{b}}\}
\end{array}
\end{equation}

\subsection{Explicit expressions}

\begin{equation}
\begin{array}{ll}
R^{(2)}_{\quad 00} = \frac{1}{2} \{ \left[h_{01,1}- \left(
\frac{h_{00} + h_{11} + h_{22} + h_{33}}{2} \right)_{,0} \right]
h_{00,0} + \left[ h_{10,0} - \left( \frac{h_{00} + h_{11} - h_{22}
- h_{33}}{2} \right)_{,1} \right] \left(  2 h_{01,0} - h_{00,1}
\right) + \\
\\
+ h_{11} \left( h_{00,11} + h_{11,00} -2 h_{01,01} \right) -
\left( h_{01,0} - h_{00,1} \right)^2  + \frac{1}{2} \left[(h_{00
,0})^2+ (h_{11 ,0})^2 - 2 (h_{01 ,0} )^2\right]+
\\
\\
+\frac{1}{2} \left[(h_{22 ,0})^2+ (h_{33 ,0})^2 + 2 (h_{23 ,0}
)^2\right]+  \left[h_{22} h_{22 ,00} + h_{33} h_{33 ,00} + 2
h_{23} h_{23 ,00}\right] \}
\end{array}
\end{equation}

\begin{equation}
\begin{array}{ll}
R^{(2)}_{\quad 11} = \frac{1}{2} \{ \left[h_{01,1}- \left(
\frac{h_{00} + h_{11} + h_{22} + h_{33}}{2} \right)_{,0} \right]
\left(  2 h_{01,1} - h_{11,0} \right) + \left[ h_{10,0} - \left(
\frac{h_{00} + h_{11} - h_{22}
- h_{33}}{2} \right)_{,1} \right] h_{11,1}  + \\
\\
+ h_{00} \left( h_{00,11} + h_{11,00} -2 h_{01,01} \right) -
\left( h_{01,1} - h_{11,0} \right)^2  + \frac{1}{2} \left[(h_{00
,1})^2+ (h_{11 ,1})^2 - 2 (h_{01 ,1} )^2\right]+
\\
\\
+\frac{1}{2} \left[(h_{22 ,1})^2+ (h_{33 ,1})^2 + 2 (h_{23 ,1}
)^2\right]+  \left[h_{22} h_{22 ,11} + h_{33} h_{33 ,11} + 2
h_{23} h_{23 ,11}\right] \}
\end{array}
\end{equation}

\begin{equation}
\begin{array}{ll}
R^{(2)}_{\quad 01} = \frac{1}{2} \{ h_{01} \left( h_{00,11} +
h_{11,00} -2 h_{01,01} \right) - \left(\frac{ h_{22} + h_{33}}{2}
\right)_{,0} h_{001} +
\left(\frac{ h_{22} + h_{33}}{2} \right)_{,1} h_{110} +\\
\\
+\frac{1}{2} \left[h_{22 ,0} h_{22 ,1} + h_{33 ,0} h_{33 ,1} + 2
h_{23 ,0} h_{23 ,1}\right]+ \left[h_{22} h_{22 ,01} + h_{33} h_{33
,01} + 2 h_{23} h_{23 ,01}\right] \}
\end{array}
\end{equation}

\begin{equation}
\begin{array}{ll}
R^{(2)}_{\quad ij} = \frac{1}{2} \{ \left[ \left( \frac{h_{00} +
h_{11} + h_{22} + h_{33}}{2} \right)_{,0} -h_{01,1}\right]
h_{ij,0}  + \left[ \left( \frac{h_{00} + h_{11} - h_{22}
- h_{33}}{2} \right)_{,1} -h_{10,0}\right] h_{ij,1}  + \\
\\
- h_{i2,0}  h_{j2,0} - h_{i3,0}  h_{j3,0} + h_{i2,1}  h_{j2,1} +
h_{i3,1}  h_{j3,1} + h_{00}  h_{ij,00} +h_{11}  h_{ij,11} - 2
h_{01}  h_{ij,01} \}
\end{array}
\end{equation}

\section{Einstein's tensor}\label{Appendix Einstein}

In this appendix we calculate the Einstein's tensor at linear and
second order of perturbations.

Using (\ref{R100}-\ref{R1ai}) and the definition (\ref{Einstein
tensor 2}), one easily obtains the formulas (\ref{linear einstein
eq new}) for the Einstein's tensor at first order.

In the same way one obtains the components of the Einstein's
tensor at second order, defined in (\ref{Einstein tensor 3}) as

\begin{subequations}

\begin{equation}\label{G200}
\begin{array}{ll}
G^{(2)0}_{\quad \,\,\, \, 0} = \frac{1}{2} \{ h_{11} B_{,11} -
h_{01} B_{,01}  + \frac{1}{2} B_{,0} \left(h_{11,0}  -2
h_{01,1}\right)  +
\frac{1}{2} B_{,1} h_{11,1} + \\
\\
+\frac{1}{2} \left[(h_{22,1})^2 - (h_{22,0})^2 +3(h_{23,1})^2 -
(h_{23,0})^2 + h_{22,0} B_{,0}+ h_{33,1} (h_{33,1}-h_{22,1})
\right]+ \\
\\
+h_{22} h_{22,11}+h_{33} h_{33,11}+2 h_{23} h_{23,11} \}
\end{array}
\end{equation}

\begin{equation}\label{G201}
\begin{array}{ll}
G^{(2)0}_{\quad \,\,\, \, 1} = \frac{1}{2} \{  h_{01} B_{,11} -
h_{00}
B_{,01} + \frac{1}{2} h_{11,0}  B_{,1}- \frac{1}{2} h_{00,1} B_{,0} +\\
\\
+\frac{1}{2} \left[h_{22,0}h_{22,1} +h_{33,0}h_{33,1} +
2h_{23,0}h_{23,1} \right]+ h_{22}h_{22,01} +h_{33}h_{33,01} +
2h_{23}h_{23,01} \}
\end{array}
\end{equation}

\begin{equation}\label{G211}
\begin{array}{ll}
G^{(2)1}_{\quad \,\,\, \, 1} = \frac{1}{2} \{ h_{00} B_{,00} -
h_{01}  B_{,01}  + \frac{1}{2} B_{,0} h_{00,0}  + \frac{1}{2}
B_{,1}  \left(h_{00,1}- 2 h_{01,0}\right)
+ \\
\\
+\frac{1}{2} \left[(h_{33,1})^2 - (h_{33,0})^2 +(h_{23,1})^2 - 3
(h_{23,0})^2 + h_{22,0} (h_{33} - h_{22})_{,0} - h_{33,1} B_{,1}  \right]+\\
\\ - h_{22} h_{22,00}-h_{33} h_{33,00}-2 h_{23} h_{23,00} \}
\end{array}
\end{equation}

\begin{equation}\label{G222}
\begin{array}{ll}
G^{(2)2}_{\quad \,\,\, \, 2} = \frac{1}{2} \{ h_{00} \left(A +
h_{33,0} \right) - h_{11} \left(A - h_{33,11} \right) +h_{01}
\left( h_{23,11} -h_{23,00}-2h_{33,01} \right) + \\
\\
+\frac{1}{2} h_{00,0}  \left(h_{11,0}+h_{33,0} -2 h_{01,1}\right)
+ \frac{1}{2} h_{00,1}  \left(h_{00,1}+h_{33,1} -  h_{11,1}\right)
+h_{01,0}  \left(h_{11,1}-h_{33,1} \right) - h_{01,1} h_{33,0}+\\
\\
+2 h_{23} \left( h_{23,11}-h_{23,00} \right) + h_{33} \left(
h_{33,11}-h_{33,00} \right)+
\\
\\
+\frac{1}{2} \left[(h_{23,1})^2 - (h_{23,0})^2 + (h_{33,1})^2 -
(h_{33,0})^2 + (h_{11,1})^2 - (h_{11,0})^2 + h_{11,0} h_{33,0}  +
h_{11,1} (h_{33,1}-h_{11,1}) \right]\}
\end{array}
\end{equation}

\begin{equation}\label{G223}
\begin{array}{ll}
G^{(2)2}_{\quad \,\,\, \, 3} = \frac{1}{2} \{-h_{00} h_{23,00} -
h_{11} h_{23,11} + 2 h_{01} h_{23,01} + h_{22}
(h_{23,00}-h_{23,11}) + h_{23} (h_{33,00}-h_{33,11}) +\\
\\
+\frac{1}{2} \left[ -h_{23,0} \left[\left(h_{00}+h_{11} -B
\right)_{,0} -2 h_{01,1} \right]  -h_{23,1}
\left[\left(h_{00}+h_{11} + B \right)_{,1} -2 h_{01,0}
\right]\right]\}
\end{array}
\end{equation}

\begin{equation}\label{G233}
\begin{array}{ll}
G^{(2)3}_{\quad \,\,\, \, 3} =\frac{1}{2} \{ h_{00} \left( A +
h_{22,00} \right) - 2 h_{01} h_{22,01} + h_{11} \left(h_{22,11 } -
A \right) +\frac{1}{2} h_{00,1} \left(h_{00,1} + h_{22,1} -
h_{11,1} \right)+\\
\\

+\frac{1}{2} h_{00,0} \left(h_{11,0} + h_{22,0} - 2 h_{01,1}
\right)

+\frac{1}{2} h_{11,0} \left(h_{22,0} - h_{11,0} -  \right)

+\frac{1}{2} h_{11,1} \left(2 h_{01,0} + h_{22,1} \right)

+\\
\\
+  h_{22} \left(h_{22,11} - h_{22,00}  \right) +h_{23}
\left(h_{23,11} - h_{23,00}  \right) +\\
\\
+

\frac{1}{2} \left[-(h_{23,0})^2+(h_{23,1})^2 +(h_{22,1})^2-
(h_{22,0})^2 - 2 h_{220} h_{01,1} -2 h_{01,0} h_{22,1} \right]

 \}
\end{array}
\end{equation}
where we have used the definitions (\ref{definition A B PSI}).

\end{subequations}

\subsection{Einstein's equations at second order }\label{Appendix Einstein euqation at second order}

The Einstein's equations at second order are given by

\begin{equation}
G^{(1)\alpha}_{\quad\,\,\, \beta} + G^{(2)\alpha}_{\quad\,\,\,
\beta} = 0
\end{equation}

One has

\begin{subequations}

\begin{equation}
\begin{array}{ll}
G^{0}_{\,\,\, \, 0} = \frac{1}{2} \{ B_{,11}+ h_{11} B_{,11} -
h_{01} B_{,01}  + \frac{1}{2} B_{,0} \left(h_{11,0}  -2
h_{01,1}\right) +
\frac{1}{2} B_{,1} h_{11,1} + \\
\\
+\frac{1}{2} \left[(h_{22,1})^2 - (h_{22,0})^2 +3(h_{23,1})^2 -
(h_{23,0})^2 + h_{22,0} B_{,0}+ h_{33,1} (h_{33,1}-h_{22,1})
\right]+ \\
\\
+h_{22} h_{22,11}+h_{33} h_{33,11}+2 h_{23} h_{23,11} \} =0
\end{array}
\end{equation}

\begin{equation}
\begin{array}{ll}
G^{0}_{\,\,\, \, 1} = \frac{1}{2} \{ B_{,01} +  h_{01} B_{,11} -
h_{00}
B_{,01} + \frac{1}{2} h_{11,0}  B_{,1}- \frac{1}{2} h_{00,1} B_{,0} +\\
\\
+\frac{1}{2} \left[h_{22,0}h_{22,1} +h_{33,0}h_{33,1} +
2h_{23,0}h_{23,1} \right]+ h_{22}h_{22,01} +h_{33}h_{33,01} +
2h_{23}h_{23,01} \}=0
\end{array}
\end{equation}

\begin{equation}
\begin{array}{ll}
G^{1}_{\,\,\, \, 1} = \frac{1}{2} \{-B_{,00}+ h_{00} B_{,00} -
h_{01} B_{,01}  + \frac{1}{2} B_{,0} h_{00,0}  + \frac{1}{2}
B_{,1} \left(h_{00,1}-h_{01,0}\right)
+ \\
\\
+\frac{1}{2} \left[(h_{33,1})^2 - (h_{33,0})^2 +(h_{23,1})^2 - 3
(h_{23,0})^2 + h_{22,0} (h_{33} - h_{22})_{,0} - h_{33,1} B_{,1}  \right]+\\
\\ - h_{22} h_{22,00}-h_{33} h_{33,00}-2 h_{23} h_{23,00} \} = 0
\end{array}
\end{equation}

\begin{equation}
\begin{array}{ll}
G^{2}_{ \,\,\, \, 2} = \frac{1}{2} \{\Box h_{22} - A - B_{00}+
B_{11}+ h_{00} \left(A + h_{33,0} \right) - h_{11} \left(A -
h_{33,11} \right) +h_{01}
\left( h_{23,11} -h_{23,00}-2h_{33,01} \right) + \\
\\
+\frac{1}{2} h_{00,0}  \left(h_{11,0}+h_{33,0} -2 h_{01,1}\right)
+ \frac{1}{2} h_{00,1}  \left(h_{00,1}+h_{33,1} -  h_{11,1}\right)
+h_{01,0}  \left(h_{11,1}-h_{33,1} \right) - h_{01,1} h_{33,0}+\\
\\
+2 h_{23} \left( h_{23,11}-h_{23,00} \right) + h_{33} \left(
h_{33,11}-h_{33,00} \right)+
\\
\\
+\frac{1}{2} \left[(h_{23,1})^2 - (h_{23,0})^2 + (h_{33,1})^2 -
(h_{33,0})^2 + (h_{11,1})^2 - (h_{11,0})^2 + h_{11,0} h_{33,0}  +
h_{11,1} (h_{33,1}-h_{11,1}) \right]\} = 0
\end{array}
\end{equation}

\begin{equation}
\begin{array}{ll}
G^{2}_{\,\,\, \, 3} = \frac{1}{2} \{\Box h_{23} -h_{00} h_{23,00}
- h_{11} h_{23,11} + 2 h_{01} h_{23,01} + h_{22}
(h_{23,00}-h_{23,11}) + h_{23} (h_{33,00}-h_{33,11}) +\\
\\
+\frac{1}{2} \left[ -h_{23,0} \left[\left(h_{00}+h_{11} -B
\right)_{,0} -2 h_{01,1} \right]  -h_{23,1}
\left[\left(h_{00}+h_{11} + B \right)_{,1} -2 h_{01,0}
\right]\right]\}=0
\end{array}
\end{equation}

\begin{equation}
\begin{array}{ll}
G^{ 3}_{ \,\,\, \, 3} =\frac{1}{2} \{ \Box h_{33} - A - B_{00}+
B_{11} + h_{00} \left( A + h_{22,00} \right) - 2 h_{01}
h_{22,01} + h_{11} \left(h_{22,11 } - A \right) +\\
\\
+\frac{1}{2} h_{00,1} \left(h_{00,1} + h_{22,1} - h_{11,1}
\right)+ \frac{1}{2} h_{00,0} \left(h_{11,0} + h_{22,0} - 2
h_{01,1} \right) +\frac{1}{2} h_{11,0} \left(h_{22,0} - h_{11,0} -
\right) +\\
\\
+\frac{1}{2} h_{11,1} \left(2 h_{01,0} + h_{22,1} \right) + h_{22}
\left(h_{22,11} - h_{22,00}  \right) +h_{23}
\left(h_{23,11} - h_{23,00}  \right) +\\
\\
+ \frac{1}{2} \left[-(h_{23,0})^2+(h_{23,1})^2 +(h_{22,1})^2-
(h_{22,0})^2 - 2 h_{220} h_{01,1} -2 h_{01,0} h_{22,1} \right]
\}=0
\end{array}
\end{equation}

and the following relations

\begin{equation}
\begin{array}{ll}
G^{2}_{ \,\,\, \, 2}-G^{3}_{\,\,\, \, 3} =\frac{1}{2} \{\Box
\left(h_{22} - h_{33} \right) + h_{00} \left(h_{33}- h_{22}
\right)_{,00} + 2 h_{01} \left(h_{22}-h_{33} \right)_{,01}
+ h_{11} \left(h_{33 } - h_{22}\right)_{,11} +\\
\\
-\frac{1}{2} \left( h_{22} - h_{33} \right)_{,0} \left(h_{00,0} +
h_{11,0} - B_{,0} -2 h_{01,1} \right) -\frac{1}{2} \left( h_{22} -
h_{33} \right)_{,1} \left(h_{00,1} + h_{11,1} + B_{,1} -2 h_{01,0}
\right)\\
\\
+ h_{22} \Box h_{22} -h_{33} \Box h_{33} \} = 0
\end{array}
\end{equation}

\begin{equation}
\begin{array}{ll}
G^{2}_{ \,\,\, \, 2}+G^{3}_{\,\,\, \, 3} =\frac{1}{2} \{- 2 A -
\Box B + h_{00,0} \left(h_{00,0}-2  h_{01,1} \right) -h_{11,1}
\left(h_{00,1}-2  h_{01,0} \right) +h_{11,0}
\left(h_{00,0}-  h_{11,0} \right) +\\
\\
+ \frac{1}{2} B_{,0} \left( h_{00,0}+  h_{11,0} - 2 h_{01,1}
\right) + \frac{1}{2} B_{,1} \left( h_{00,1}+ h_{11,1} - 2
h_{01,0} \right) +h_{00} \left(B_{,00} + 2 A \right) +h_{11}
\left(B_{,11} -2 A
\right)\\
\\
 -2h_{01} B_{,01} - h_{22} \Box h_{22} - h_{33} \Box h_{33} -2 h_{23} \Box h_{23}
+(h_{00,1})^2 -(h_{00,0})^2 +(h_{23,1})^2-(h_{23,0})^2 +\\
\\
+ \frac{1}{2} \left[(h_{22,1})^2 -(h_{22,0})^2
+(h_{33,1})^2-(h_{33,0})^2  \right] \}=\\
\\
=\frac{1}{2} \{- 2 A - \Box B + h_{00,0} \left(h_{00,0}-2 h_{01,1}
\right) -h_{11,1} \left(h_{00,1}-2  h_{01,0} \right) +h_{11,0}
\left(h_{00,0}-  h_{11,0} \right) +\\
\\
+ \frac{1}{2} B_{,0} \left( h_{00,0}+  h_{11,0} - 2 h_{01,1}
\right) + \frac{1}{2} B_{,1} \left( h_{00,1}+ h_{11,1} - 2
h_{01,0} \right) +h_{00} \left(B_{,00} + 2 A \right) +h_{11}
\left(B_{,11} -2 A
\right)\\
\\
 -2h_{01} B_{,01} - h_{22} \Box h_{22} - h_{33} \Box h_{33} -2 h_{23} \Box h_{23}
+(h_{00,1})^2 -(h_{00,0})^2 +(h_{23,1})^2-(h_{23,0})^2 +\\
\\
+ \frac{1}{4} \left[(B_{,1})^2 -(B_{,0})^2
+(\psi_{,1})^2-(\psi_{,0})^2  \right] \} = 0
\end{array}
\end{equation}

\begin{equation}
\begin{array}{ll}
G^{0}_{\,\,\, \, 0}+ G^{1}_{\,\,\, \, 1} = \frac{1}{2} \{- \Box B+
B_{,00}  h_{00} +B_{,11}  h_{11}  - 2 B_{,01} h_{01}  +
\frac{1}{2} B_{,0} \left(h_{00,0}+ h_{11,0}  - 2 h_{01,1} \right) + \\
\\
+ \frac{1}{2} B_{,1} \left(h_{00,1}+ h_{11,1}  - 2 h_{01,0} \right) -h_{22} \Box h_{22}-h_{33} \Box h_{33}-2 h_{23} \Box h_{23}\\
\\
+\frac{1}{2} \left[\left( \psi_{,1} \right)^2 - \left( \psi_{,0}
\right)^2\right] + 2 \left[ \left( h_{23,1} \right)^2 - \left(
h_{23,0} \right)^2 \right] \} = 0
\end{array}
\end{equation}

\begin{equation}
\begin{array}{ll}
G^{2}_{\,\,\, \, 2}+ G^{3}_{\,\,\, \, 3}  - \left(G^{0}_{\,\,\, \,
0}+ G^{1}_{\,\,\, \, 1}\right) =\\
\\
=\frac{1}{2} \{-2 A + (h_{00,1})^2 -(h_{11,0})^2  +
 (h_{23,0})^2 -(h_{23,1})^2  + \frac{1}{4}
\left[(B_{,1})^2 -(B_{,0})^2 +(\psi_{,0})^2 -(\psi_{,1})^2
\right]+\\
\\
+ 2 \left(h_{00}-h_{11} \right) A  + h_{00,0} \left(h_{11,0}-2
h_{01,1} \right)- h_{11,1} \left(h_{00,1}-2 h_{01,0} \right)\} = 0
\end{array}
\end{equation}

\end{subequations}

\section{Solution of the equation (\ref{solh00})}\label{Appendix prototipe equation}

The equation (\ref{solh00}), which governs the evolution of
$h_{11}$,   is of the type

\begin{equation}\label{equation}
\Box U = U_x^2 - U_t^2
\end{equation}

This equation is obtained by the following lagrangian

\begin{equation}\label{lagrangian}
L =  \left( U_t^2 - U_x^2 \right) \exp(2 U)
\end{equation}
and the corresponding hamiltonian is

\begin{equation}\label{hamiltonian}
H =  \left( U_t^2 + U_x^2 \right) \exp(2 U)
\end{equation}
Equation (\ref{equation}) is solved by plane waves of the type $U
= U(t-x)$ or $U = U(t+x)$, but a linear combination of these two
solutions is not a solution itself.

The general solution of (\ref{equation}) is obtained by the simple
observation that, defining

\begin{equation}\label{redefinition}
\psi \equiv \exp\left(U\right)
\end{equation}
the lagrangian (\ref{lagrangian}) becomes

\begin{equation}\label{lagrangian2}
L =  (\psi_t)^2 - (\psi_x)^2
\end{equation}
which gives the standard wave equation
\begin{equation}\label{lagrangian2}
\Box \psi =0
\end{equation}
whose general solution is
\begin{equation}\label{exact solution 1}
\psi(x^0,x^1) = A(x^0+x^1) + B(x^0-x^1)
\end{equation}

Therefore the solution of (\ref{equation}) is

\begin{equation}\label{exact solution 1}
U(x^0,x^1) = \ln\left(A(x^0+x^1) + B(x^0-x^1) \right)
\end{equation}

Since, in the case of equation (\ref{eeeqh00}), one has $h_{11}
\sim \epsilon$, one has

\begin{equation}\label{exact solution 1}
h_{11}(x^0,x^1) = \ln\left(1 + \alpha(x^0+x^1) + \beta(x^0-x^1)
\right)
\end{equation}
with $\alpha(x^0+x^1) \sim \beta(x^0-x^1) \sim \epsilon$.

\end{document}